\documentclass[12pt]{article}
\usepackage{amsmath,epsfig,graphicx,amssymb}

\newcommand{\beq}{\begin{equation}}
\newcommand{\eeq}{\end{equation}}
\newcommand{\ben}{\begin{eqnarray}}
\newcommand{\een}{\end{eqnarray}}
\newcommand{\bes}{\begin{subequations}}
\newcommand{\ees}{\end{subequations}}
\newcommand{\bFig}{\begin{figure}}
\newcommand{\eFig}{\end{figure}}

\begin{document}

\title{On Entangled Multi-particle Systems in Bohmian
Theory}
\author{Partha Ghose \\
Centre for Natural Sciences and Philosophy, \\ 1/AF Bidhan Nagar,
Kolkata, 700 064, India\\ and\\Centre for Philosophy and Foundations of Science,\\Darshan Sadan, E-36 Panchshila Park,
New Delhi 110017, India}

\maketitle

\begin{abstract}

Arguments are presented to show that in the case of entangled systems there are certain difficulties in implementing the usual Bohmian interpretation of the wave function in a straightforward manner. Specific examples are given.
\end{abstract}

The three basic prescriptions of standard
Bohmian quantum theory \cite{bohm} are:

(i) take the wave function $\psi$ to be a solution of the
Schr\"{o}dinger equation,

(ii) impose the guidance condition ${\bf p} = m\, d{\bf x}/dt =
{\bf \nabla}\, S$ where $S$ is the phase of the wave function
$\psi = R\, {\rm exp}(i S/\hbar)$, and

(iii) choose the particle distribution $P_{[t_0]}$ at some {\it arbitrary} time $t_0$ (the initial time) such that
$P_{[t_0]} = \vert \psi \vert_{[t_0]}^2 = R^2_{[t_0]}$.
This is known in the literature as the `quantum equilibrium hypothesis' (QEH).
\noindent Given the prescriptions (i) through (iii), one can prove
complete equivalence between this theory and standard quantum
mechanics by using the continuity equation for $R^2$ to show that
$P_{[t]} = R^2_{[t]}$ for all subsequent times.

While these prescriptions are self-consistent and work for single
particle and factorizable many-particle systems, it turns out that
non-factorizable multi-particle systems sometimes present certain difficulties. This is principally because the
velocity equations for such systems are non-separable and imply constraints on particle positions which make implementation of prescrition (iii) problematic. Although no general
method is known yet to calculate the trajectories for such systems exactly, 
analogous to the case of the Hamilton-Jacobi theory of many-particle systems in classical
mechanics \cite{goldstein}, we give two examples of systems in which the constraint can be calculated. 

Let us consider an $N$-particle entangled
system described by the wave function \beq \psi(x_1, x_2,..., x_N,t)=
R (x_1, x_2, ..., x_N,t)e^{\frac{i}{\hbar}S(x_1, x_2, ...,
x_N,t)}\label{wavfn1}\eeq The guidance conditions which define the
velocities are \ben v_i(x_1, x_2,..., x_N, t) = \frac{d x_i}{d t} &=&
\frac{1}{m}\frac{\partial S(x_1, x_2,..., x_N, t)}{\partial
x_i}\nonumber\\&=& f_i (x_1, x_2,..., x_N, t)\,\,\,\,\,\,\,\,i=
1,2,...,N \een The coordinates $x_i$ are not separable, and hence
the velocity equations are not separable for entangled states, for
had they been so, the wave function (\ref{wavfn1}) would have been
factorizable which by assumption it is not. Since the equations
cannot be separated, no general method is known to solve them exactly to
find the trajectories. 

Consider now the example of a two-particle non-factorizable wave
function \beq \psi_{[a,b,p]}(x_1, x_2, t) = \frac{1}{\sqrt{L}}[a\, e^{ip (x_1 -
x_2)/\hbar} + b\, e^{-ip (x_1 - x_2)/\hbar}]\,e^{- \frac{i
Et}{\hbar}}\label{wavfn} \eeq where $x_1$ and $x_2$ are the coordinates of the two distinguishable particles of the same mass $m$ and momentum $p$, $\sqrt{L}$ is a normalization
constant, and $a$ and $b$ are real parameters with $a^2 + b^2 = 1$ and $a \neq b$. Different choices of the parameters $a$, $b$ and $p$ correspond to different wave functions. 
We assume box normalization with $- (2N + 1)\frac{\pi}{2} < p\, (x_1
- x_2)/\hbar < (2N + 1)\frac{\pi}{2}$, $N$ being a finite but
sufficiently large integer. Then $L = (2N + 1)\pi (a^2 + b^2) \hbar
\,/p = (2N + 1)\pi \hbar
\,/p $ which is a multiple of the de Broglie wavelength of the particles, the only natural scale in the theory. This is a solution of the Schr\"{o}dinger equation provided
$E = p^2/m$. To keep the notation simple, we shall henceforth drop the suffix $[a,b,p]$ from the wave function. The position probability density is given by \beq
R^2(x_1, x_2, t)=\vert \psi(x_1, x_2, t)\vert^2 = \frac{1}{L}[1 + 2
ab\,\, {\rm cos}\, 2p\, (x_1 - x_2)/\hbar]\label{probden}\eeq The
important points about this probability density, as far as this
paper is concerned, are that (a) it is stationary, (b) it depends
only on $(x_1 - x_2)$ and (c) $(x_1 - x_2)$ can take all values in
the support of the wave function {\it at all times}.

The phase of the wave function (\ref{wavfn}) is \ben S(x_1, x_2, t)
&=& \hbar\, {\rm arc\, tan}\, [(\frac{a - b}{a + b})\,{\rm
tan}\,p\,(x_1 - x_2)/\hbar]\, - E t + \eta \hbar \nonumber\\&=&
\hbar\, {\rm arc\,tan}\,\theta - E t + \eta \hbar\label{phase1}\een
where \beq \theta = (\frac{a - b}{a + b})\,{\rm tan}\, p\,(x_1 -
x_2)/\hbar\label{phase2}\eeq and $\eta = (n + 1) \pi$ for $(2n +
1)\pi/2< {\rm arc\,tan}\, \theta < (2n + 1)\pi/2 + \pi$, $\eta = -
n\pi$ for $-(2n + 1)\pi/2< {\rm arc\,tan}\,\theta < -(2n + 1)\pi/2 +
\pi$, $n = 0, 1, 2,,...$\,. In order to have a continuous and
single-valued wave function, we choose $\eta = 0$, i.e., the
principal branch of the function $S(x_1, x_2, t)$ which lies between
$\pm \pi/2$. The Bohmian guidance conditions are therefore \ben v_1
&=& \frac{d x_1}{d t} =
\frac{1}{m}\frac{\partial S(x_1, x_2, t)}{\partial x_1}\nonumber\\
&=& \frac{p}{m[1 + \theta^2]}(\frac{a - b}{a + b})\,{\rm sec}^2
\frac{p}{\hbar}(x_1 - x_2)\nonumber\\&=& \frac{p}{m}\frac{(a - b)/(a + b)}{{\rm cos}^2
p\,(x_1 - x_2)/\hbar + [(a - b)/(a + b)]^2{\rm sin}^2 p\,(x_1 -
x_2)/\hbar}\label{guid1}\\v_2 &=& \frac{d x_2}{d t} =
\frac{1}{m}\frac{\partial S(x_1, x_2, t)}{\partial x_2}\nonumber\\
&=& - \frac{p}{m}\frac{(a - b)/(a + b)}{{\rm cos}^2 p\,(x_1 -
x_2)/\hbar + [(a - b)/(a + b)]^2{\rm sin}^2 p\,(x_1 -
x_2)/\hbar}\label{guid2}\een The coordinates $x_i$ are non-separable
and so are these equations, i.e., the velocity of each particle
depends not only on its own position but also on the position of the
other. Note, however, that (\ref{guid1}) and (\ref{guid2}) imply \beq
v_1 + v_2 = 0\label{cm} \eeq which shows that the centre-of-mass of
the particles is stationary.

It also follows from (\ref{guid1}) and (\ref{guid2}) that \beq
\frac{d (x_1 - x_2)}{d t} = \frac{2p}{m}\frac{(a - b)/(a + b)}{{\rm
cos}^2 p\,(x_1 - x_2)/\hbar + [(a - b)/(a + b)]^2{\rm sin}^2 p\,(x_1
- x_2)/\hbar}\label{guid3} \eeq This equation for $(x_1 - x_2)$ is
integrable. Using $a^2 + b^2 = 1$, the solution is \beq \frac{1}{2(a^2 -
b^2)}(x_1 - x_2) + \frac{\hbar}{p}\frac{a b}{(a^2 - b^2)}\,{\rm
sin}\,2 p\,(x_1 - x_2)/\hbar = \frac{2p}{m}t +
\beta\label{soln1}\eeq where $\beta$ is an arbitrary constant of
integration. This is a constraint on the particle positions, as we will now show. Since the phase $S (x_1, x_2,t)$ (Eqn. (5)) is well defined for $x_1 - x_2 = 0$ and the wave function (\ref{wavfn}) does not vanish at this point, this equality must hold at some time. Let this time be $t_0$. Then $\beta = - 2pt_0/m$ and hence
\beq \frac{1}{2(a^2 -
b^2)}(x_1 - x_2) + \frac{\hbar}{p}\frac{a b}{(a^2 - b^2)}\,{\rm
sin}\,2 p\,(x_1 - x_2)/\hbar = \frac{2p}{m}(t - t_0)\label{soln2}\eeq It is straightforward to see from this that $x_1 - x_2 = 0$ is {\it the only solution} at $t = t_0$ {\it provided} $4 ab < 1$. Hence, given $4 ab < 1$, {\it every pair of particles} in the ensemble at $t = t_0$ must satisfy this constraint. If there are other times $t_0^\prime, t_0^{\prime\prime}$, etc. at which also the constraint (\ref{soln1}) holds, every pair in the ensemble must meet at such times too provided $4 ab < 1$. Therefore, given the range of parameters $4 ab < 1$, the particle distribution in the Bohmian theory cannot be chosen to match the quantum mechanical distribution (\ref{probden}) at these times -- one must avoid these times to invoke QEH. This makes a Bohmian interpretation of these wave functions problematic. No such problem arises with single-particle and factorizable many-particle wave functions.

Another example of a constrained system is the following. Consider
the two-particle wave function
\begin{equation} \psi (r_{1}, r_{2}, t) = \frac{1}{N}[ \frac{e^{i
k (r_{1A} + r_{2B})}}{r_{1A} r_{2B}} + \frac{e^{i k (r_{1B} +
r_{2A})}}{r_{1B} r_{2A}}]\,e^{-\frac{i}{\hbar} E t}\
\label{wavfn2}
\end{equation}
where $N$ is a normalization factor and
\begin{eqnarray}
r_{1A} &=& \sqrt{x_1^2 + (y_1 - a)^2 + z_1^2}\,\,\,\,\,\, r_{2B}
=\sqrt{x_2^2 + (y_2 + a)^2 + z_2^2}\label{coord1}\\ r_{1B} &=&
\sqrt{x_1^2 + (y_1 + a)^2 + z_1^2}\,\,\,\,\,\,r_{2A} = \sqrt{x_2^2
+ (y_2 - a)^2 + z_2^2}\label{coord2}\end{eqnarray} where the first
index $i$ (1,2) in $r_{ij}$ denotes the particle and the second
index $j$ denotes a point-like slit $A$ of co-ordinates (0,a,0) or
a point-like slit B of co-ordinates (0,-a,0) which are sources of
the two spherical waves in the $x \geq 0$ space. This wave
function is normalizable in a finite volume, analogous to the
plane wave case. This wave function is separately symmetric under
reflection in the $x$ axis ($y_i \rightarrow -y_i$) {\it and} the
interchange of the two particles $1 \leftrightarrow 2$.

The phase $S$ of the wave function (\ref{wavfn2}) is (in an
obvious notation)
\begin{eqnarray}
S &=& \hbar\, {\rm arctan} \frac{r_{1B}r_{2A}\, {\rm sin}k (r_{1A}
+ r_{2B}) + r_{1A}r_{2B}\, {\rm sin}k (r_{1B} +
r_{2A})}{r_{1B}r_{2A}\, {\rm cos}k (r_{1A} + r_{2B}) +
r_{1A}r_{2B}\, {\rm cos}k (r_{1B} + r_{2A})}\nonumber\\&=&
\hbar\, {\rm arctan}\frac{N}{D} \label{eq:phase}
\end{eqnarray}
with
\begin{eqnarray}
N &=& r_{1B}r_{2A}\, {\rm sin}k (r_{1A}
+ r_{2B}) + r_{1A}r_{2B}\, {\rm sin}k (r_{1B} +
r_{2A})\\
D &=& r_{1B}r_{2A}\, {\rm cos}k (r_{1A} + r_{2B}) +
r_{1A}r_{2B}\, {\rm cos}k (r_{1B} + r_{2A})
\end{eqnarray}
The Cartesian components of the Bohmian velocities of the two
particles can be computed from $S$ using
\begin{eqnarray}
v_{x_1} &=& \frac{d x_1}{d t} = \frac{1}{m}\frac{\partial
S}{\partial x_1} = \frac{1}{m}(\frac{\partial S}{\partial
r_{1A}}\frac{\partial r_{1A}}{\partial x_1} + \frac{\partial
S}{\partial r_{1B}}\frac{\partial r_{1B}}{\partial
x_1})\label{eq:vx1}\\v_{y_1} &=& \frac{d y_1}{d t}
=\frac{1}{m}\frac{\partial S}{\partial y_1} =
\frac{1}{m}(\frac{\partial S}{\partial r_{1A}}\frac{\partial
r_{1A}}{\partial y_1} + \frac{\partial S}{\partial
r_{1B}}\frac{\partial r_{1B}}{\partial
y_1})\label{eq:vy1}\\v_{z_1} &=& \frac{d z_1}{d t}
=\frac{1}{m}\frac{\partial S}{\partial z_1} =
\frac{1}{m}(\frac{\partial S}{\partial r_{1A}}\frac{\partial
r_{1A}}{\partial z_1} + \frac{\partial S}{\partial
r_{1B}}\frac{\partial r_{1B}}{\partial
z_1})\label{eq:vz1}\\v_{x_2} &=& \frac{d x_2}{d t} =
\frac{1}{m}\frac{\partial S}{\partial x_2} =
\frac{1}{m}(\frac{\partial S}{\partial r_{2A}}\frac{\partial
r_{2A}}{\partial x_2} + \frac{\partial S}{\partial
r_{2B}}\frac{\partial r_{2B}}{\partial
x_2})\label{eq:vx2}\\v_{y_2} &=& \frac{d y_2}{d t}
=\frac{1}{m}\frac{\partial S}{\partial y_2} =
\frac{1}{m}(\frac{\partial S}{\partial r_{2A}}\frac{\partial
r_{2A}}{\partial y_2} + \frac{\partial S}{\partial
r_{2B}}\frac{\partial r_{2B}}{\partial y_2})\label{eq:vy2}\\
v_{z_2} &=& \frac{d z_2}{d t} =\frac{1}{m}\frac{\partial
S}{\partial z_2} = \frac{1}{m}(\frac{\partial S}{\partial
r_{2A}}\frac{\partial r_{2A}}{\partial z_2} + \frac{\partial
S}{\partial r_{2B}}\frac{\partial r_{2B}}{\partial
z_2})\label{eq:vz2}\\
\end{eqnarray}
where
\begin{eqnarray}
\frac{\partial S}{\partial r_{1A} }&=& \hbar [1 + N^2/D^2]^{-1}
[\frac{k r_{1B}r_{2A}\, {\rm cos}k (r_{1A} + r_{2B})+ r_{2B}\,
{\rm sin}k (r_{1B} + r_{2A}) }{D}\nonumber\\ &-& \frac{N}{D^2}(- k
r_{1B}r_{2A}\, {\rm sin}k (r_{1A} + r_{2B}) +
r_{2B}\, {\rm cos}k (r_{1B} + r_{2A}))]\label{eq:diff1}\\
\frac{\partial S}{\partial r_{2B}}&=& \hbar [1 + N^2/D^2]^{-1}
[\frac{k r_{1B}r_{2A}\, {\rm cos}k (r_{1A} + r_{2B})+ r_{1A}\,
{\rm sin}k (r_{1B} + r_{2A}) }{D}\nonumber\\ &-& \frac{N}{D^2}(-
k r_{1B}r_{2A}\, {\rm sin}k (r_{1A} + r_{2B}) + r_{1A}\, {\rm
cos}k (r_{1B} + r_{2A}))]\label{eq:diff2}
\end{eqnarray}
The expressions for $\partial S/\partial r_{1B}$ and $\partial
S/\partial r_{2A}$ are easily obtained by the replacements $A
\leftrightarrow B$ in the above expressions. These show that the
differential equations for the velocities of the two particles are
non-separable. As we have seen, this is a general feature of
many-particle entangled systems in Bohmian theory.

It is clear from the velocity equations (\ref{eq:vx1}) through
(\ref{eq:diff2}) that the equation for each particle can be
written solely in terms of its own coordinates provided
\begin{equation}
r_{1A} = r_{2B}\,\,\,\,\,\, {\rm and}\,\,\,\, r_{1B} =
r_{2A}\label{R}
\end{equation}
These are therefore `integrability conditions' for the velocity
equations, or equivalently, constraints that the trajectories must
satisfy at all times. Notice that no assumption has been made about the initial positions of the particles in arriving at these conditions --{\it they are independent of initial conditions}. The existence and properties of other trajectories, if they exist, remain conjectural.

These two examples clearly demonstrate that there {\it are} multiparticle entangled wave functions like (\ref{wavfn}) and  (\ref{wavfn2}) in standard quantum mechanics for which there is no straightforward de Broglie-Bohm interpretation. 

Earlier attempts by Ghose \cite{ghose} and others \cite{Golshani} to show incompatibility between standard quantum mechanics and Bohmian theory were criticised mainly on the ground that the initial distributions assumed in these papers were incompatible with QEH which is an integral part of Bohmian theory. For a full account of the controversies this kind of criticism generated, see Struyve and De Baere \cite{struyve} and references therein.

An experiment was also performed by Brida {\it et al} \cite{Brida} which claimed to simulate the kind of constrained system considered in the second example given above. Their claim that the observed results were incompatible with Bohmian theory was criticised for the same reason, namely non-compliance with QEH, this time by Oriols \cite{Oriols}. 

The analyses of the two constrained systems considered above in this paper show that QEH cannot be invoked for wave functions of entangled multiparticle systems in general. Wave functions (\ref{wavfn}) and  (\ref{wavfn2}) are examples. In the first example (\ref{wavfn}) the initial conditions cannot be chosen to fit the quantum mechanical distribution at arbitrary times, and in the second example (\ref{wavfn2}) the particle distribution is incompatible with QEH at all times, independent of initial conditions.

\end{document}